# Quartz Whispering-Gallery-Mode Resonator with Microfluidic Chip as Sensor for Permittivity Measurement of Liquids

Alexey I. Gubin, *Member, IEEE,* Irina A. Protsenko, *Member, IEEE,* Alexander A. Barannik, *Member, IEEE,* Svetlana Vitusevich, Alexandr A. Lavrinovich, Nickolay T. Cherpak, *Senior Member, IEEE*

*Abstract*—Studies of biological solutions require high measurement accuracy, the ability to detect low changes in substance concentration, and small amounts of the liquid under test. A microwave complex permittivity measurement technique based on a high-quality-factor whispering-gallery-mode resonator with a microfluidic chip allows small amounts of dielectric liquids to be investigated with high accuracy. An existing technique based on a sapphire resonator does not provide for a low-concentration detection limit for substances with low molecular weight. Here we present an advanced technique based on a quartz dielectric resonator. The detection limit obtained for a glucose in water solution was found to be about one order of magnitude lower for the quartz resonator cell than for the sapphire-resonator-based measurement cell. The limit is about or lower than the concentration of glucose in human blood. This fact means that the technique can be used as the sensor for investigations of biological solutions in the microwave range. A study of glucose, lactalbumin, and bovine serum albumin was successfully performed in the Ka band using the developed sensor.

*Index Terms*—dielectric liquids, microfluidics, microwave measurement, permittivity

## I. Introduction

THE investigation of liquids and solutions is important in many branches of life and science. Different liquids have definite frequency dependencies based on their characteristics; therefore, it is important to study liquid permittivity both to identify liquids and to investigate different processes in them [1]. Microwave measurement techniques utilize some volumes filled with liquid under test (LUT). Resonant approaches based on the use of different resonant structures allow investigation at certain discrete frequencies only, in comparison to non-resonant approaches developed on the basis of different parts of transmission lines. However, the resonant approaches are more sensitive [2],[3]. A technique based on a sapphire whispering-gallery-mode (WGM)

A. I. Gubin, I. A Protsenko, A. A. Barannik, A. A. Lavrinovich, and N. T. Cherpak are with the Department of solid-state radiophysics, O. Ya. Usikov Institute for Radiophysics and Electronics, NAS of Ukraine, 61085 Kharkiv, Ukraine (e-mail: gubin@ire.kharkov.ua).

S. Vitusevich is with the Institute of Complex Systems – Bioelectronics (ICS-8), Forschungszentrum Jülich, 52425 Jülich, Germany (corresponding author; e-mail: s.vitusevich@fz-juelich.de).

A. I. Gubin gratefully acknowledges a research grant from the German Academic Exchange Service (DAAD).

The authors would like to acknowledge the German Research Foundation (DFG), project VI 456/3-1, for partial financial support.

resonator covered by a plastic layer with a microfluidic channel delivers high accuracy (about 0.7% for the imaginary part of the permittivity and 1.5% for the real part) [4],[5] compared to other microwave methods. The high accuracy and sensitivity of the method is conditioned by the high quality factor of WGM resonators. Such a technique has a number of other advantages – controllability of mode coupling, possibility to investigate small liquid amounts, and short measurement times (in the ms range). These advantages open up prospects for the technique in the investigation, testing, and monitoring of biochemical liquids.

One of the important parameters of the measurement technique for the investigation of biochemical liquids is the detection limit of the particular substance. The detection limit of the substance's concentration in a solution depends on the substance's molecular weight. For bovine serum albumin solution (i.e. a substance with high molecular weight), the concentration detection limit of the technique based on the sapphire WGM resonator is equal to about 50 μmol/l. However, for glucose in water solution, it is equal to about 30 mmol/l. The concentration of low-molecular substances in real biological solutions is lower. The concentration of glucose in human blood varies in the range from about 3 to 7 mmol/l. In this respect, the concentration detection limit must be enhanced in order to use the technique for real biological applications.

It is well known that sapphire has a strong permittivity dependence on temperature [6]. To achieve maximal accuracy, the sapphire-resonator-based measurement cell was placed in a temperature-isolated box. The temperature in the box was stabilized with an accuracy better than 0.01°C. Nevertheless, the main source of errors in the setup was temperature-related changes in the sapphire resonator parameters. Due to technical reasons, it was not possible to enhance the temperature stabilization simply. Quartz has both a lower thermal expansion coefficient and a lower temperature dependence of permittivity in comparison with sapphire [7]. Using a quartz resonator instead of a sapphire resonator allows the temperature dependence of the resonance frequency to be decreased about eightfold [8]. At the other hand using of the quartz resonator based sensor is important in investigation of the temperature LUT parameters dependencies, for example liquor cerebrospinalis investigation in the range of 30-80°C.

Here we present results demonstrating the utilization of a quartz dielectric resonator with a microfluidic chip as the basis for an approach to determine the complex permittivity of LUT. The modeling of the resonant structure was performed using COMSOL Multiphysics [9] commercial software. The measurement cell parameters were optimized for the measurement of aqueous solutions and comparison with the sapphire resonator measurement cell technique. A suitable resonance mode was also chosen for comparison; however, it is possible to perform measurements using several different modes (Section III.A). The reliability of the calculation was checked by comparing the measured and simulated results for a number of reference liquids with known characteristics (Section III.B). The resonance frequency shift and change in inverse quality factor of the measurement cell with glucose in water solutions of different concentrations were measured (Section III.C). The glucose in water permittivity obtained using a calibration procedure based on a nomogram chart (Section III.D) was compared with the results obtained using the sapphire WGM-resonator-based measurement cell [4] (Section III.E). As well as the complex permittivity of glucose, the permittivity values of bovine serum albumin and lactalbumin aqueous solutions of various concentrations were measured (section III.F). The minimum concentration detection limit of the quartz-resonator-based measurement cell was determined for substances with low and high molecular weight.

## II. EXPERIMENTAL AND MODELING DETAILS

### A. Experimental technique

The quartz resonator measurement cell was designed using the same main principles that were applied for the fabrication of the sapphire resonator measurement cell [4]. The resonator dimensions were as follows: diameter – 25.00 mm, height – 4.93 mm, inner hole diameter – 3.05 mm (Fig.1). Quartz is an anisotropic material with the permittivity in Ka-band in perpendicular and parallel directions to the crystal optical axis equal to $\varepsilon_\perp = 4.44(1-1.1\times10^{-5})$ and $\varepsilon_{//} = 4.63(1-8\times10^{-6})$, respectively. Anisotropic axis is parallel to the axis of resonator cylindrical disk.

The quartz resonator dimensions are larger in comparison to the sapphire resonator, since the quartz permittivity is lower. A pair of Teflon mirror waveguides was used to couple the resonator with microwave transmission lines. A microfluidic chip was fabricated from ZEONOR 1420 low-loss plastic by means of thermocompression. The suitability of such a material for this purpose was checked in the sapphire-resonator-based measurement cell [4]. The microfluidic channel of 0.21 mm diameter was placed in the middle of the chip height. The chip dimensions were 30x40 mm and the thickness was 0.73 mm. The distance of the channel from the resonator axis in the sapphire-resonator-based measurement cell was tuned to achieve maximal sensitivity for the investigation of water solutions [4]. In this respect and for comparison with the measurement cell, the distance of the channel from the resonator axis was determined by selection of quality factor value for the quartz and sapphire measurement cells with a water-filled microfluidic channel to be equal. The quality factor for the quartz resonator based sensor and for sapphire one was measured to be about 3665 and 3715, respectively. The distance thus obtained was about 8 mm. The metal tubes were sealed at the edges of the chip to be filled with the LUT. The minimal volume of LUT required for measurements was about 1 µL. The amount of liquid can be further decreased by reducing the microfluidic channel diameter and placing the channel closer to the resonator surface. The measurement cell was placed in a temperature-isolated box with a temperature stabilization system delivering an accuracy of better than 0.01°C. The silicon tubes were used to fill the microfluidic chip with the LUT outside the isolated box.

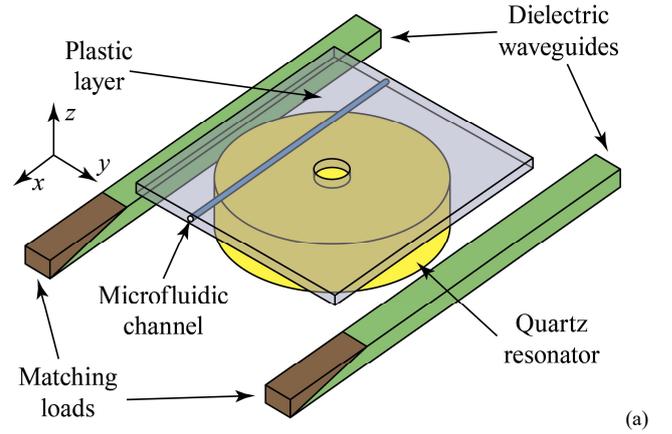
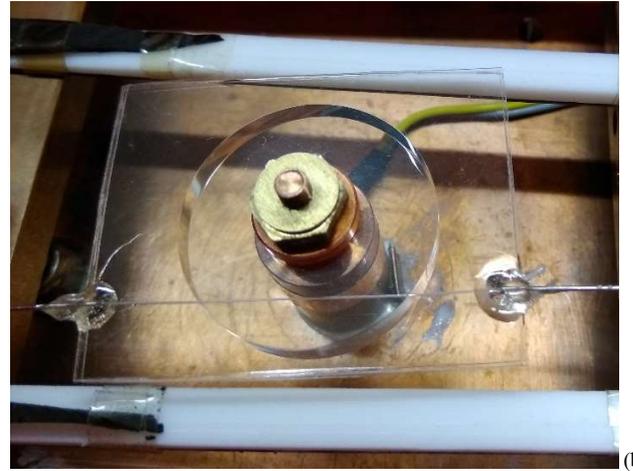

Fig. 1. (a): Schematic of the measurement cell consisting of a quartz cylindrical disk (WGM resonator) covered with a microfluidic chip (ZEONOR 1420 plastic layer with a microfluidic channel and stainless steel tubes for filling the liquid under test). (b): Photograph of the measurement cell.

### B. Simulation details

COMSOL Multiphysics [9] was used to model the quartz-resonator-based measurement cell. The reliability of using commercial software for such a simulation was tested by modeling the sapphire WGM-resonator-based measurement cell [4].

The simulations were performed using a radio-frequency (RF) module. An eigenvalue solver was used to obtain the resonance frequency and quality factor of the resonator with microfluidic chip.

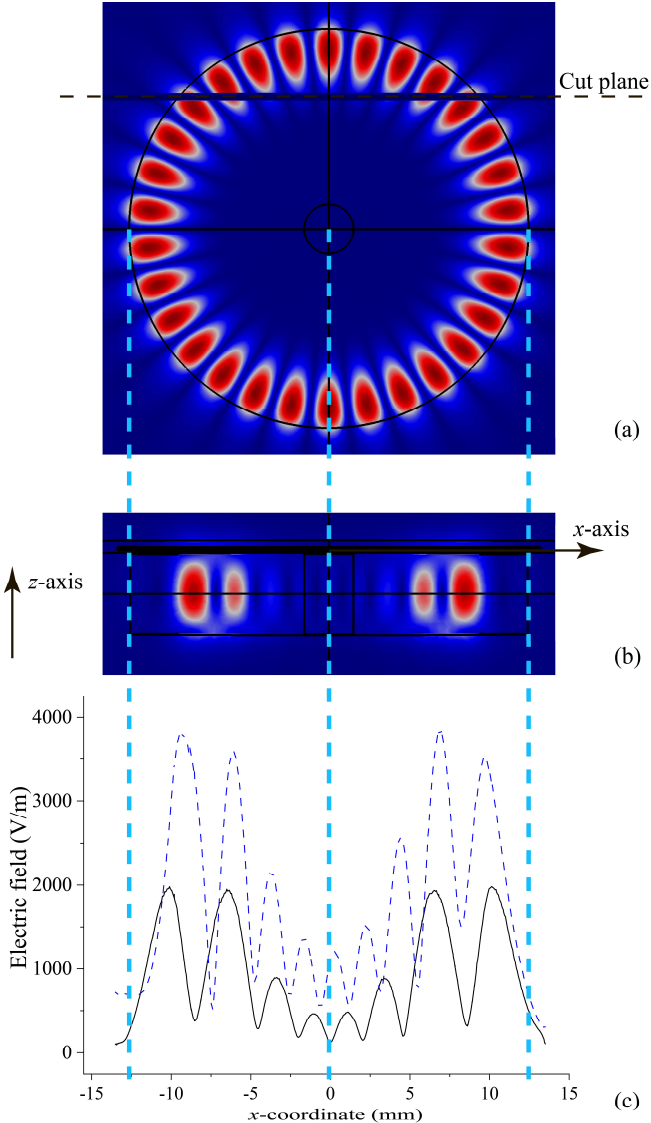

Fig. 2. Distribution of electric field $E_z$ component of $HE_{15\,1\,1}$ mode in the resonator (a): in the *x-y* plane; (b): in the *z-x* plane of cross section of the resonator along the line of the microfluidic channel; and (c): the $E_Z$-component values as a function of *x*-coordinate in the channel filled with water (solid lines) and air (dash lines).

The diameter of the microfluidic channel and its position were fitted by calculation within the measurement accuracy of the chip geometrical dimension measurement. The fitting was performed in a similar way as for the sapphire-resonator-based measurement cell [4]. In such a way it was established that diameter of the channel is 0.215 mm and the distance between the resonator axis and the channel is 7.9 mm.

Fig.2 demonstrates the field pattern in the resonator with the microfluidic chip. Fig 2a shows the distribution of normalized $E_z$-component. The dashed line in this figure marks cut plane parallel to the resonator axis through the microfluidic channel. The cross section obtained in such a way reflects the field pattern in the *z-x* plane (Fig.2b). Microfluidic channel filled with water due to the small volume of the channel has similar field distribution as patterns corresponding to air-filled channel. However, it should be emphasized that in both cases the electric intensity in the area of channel is different. Fig.2c shows the normalized values of $E_z$-component along the *x*-axis (see Fig 2 a) for these cases.

High values of water permittivity causes stronger field penetration into the channel (see solid lines in the Fig.2c), but additional losses cause the field attenuation in the channel area in comparison with the field in air-filled channel (dash lines in the Fig.2c).

The model made it possible to observe the mode splitting caused by the presence of the microfluidic channel. The resonance frequencies of the two split modes were so close to each other (the difference was in the range of 0.1 to 0.3 MHz) that it was impossible to separate them in the experiment. Therefore, in order to compare them to experimental results, average values were calculated.

III. RESULTS AND DISCUSSION

The resonance frequency and quality factor can be obtained in a number of ways. An 8722C vector network analyzer, which was used for measurements in the experiment, determines the above-mentioned parameters by estimating the resonance half-width at a level of -3 dB. Fitting using a Lorentzian function gives a more accurate result, but only in an ideal scenario. In practice, recovery of the coupled mode parameters is the most accurate method [11], and this was tested during the measurements. The method is similar to fitting with a Lorentzian function, but is based on fitting the amplitude frequency response with the sum of the complex mode amplitudes. All resonance parameters in the work were obtained in this way.

*A. Bare quartz resonator characteristics*

The spectral characteristics of a bare quartz resonator were investigated to find the optimal measurement mode. Table 1 shows the experimentally obtained resonance frequencies and quality factors together with the results of the simulation. The $HE_{15\,1\,1}$ mode was found to be optimal in terms of achieving a sufficiently high quality factor. On the other hand, the frequency was close to the $HE_{12\,1\,1}$ mode (quality factor is about 40000) in the sapphire resonator [4], which is important for comparing the techniques. Modes with a high quality factor are suitable for measuring LUT permittivity. In the case of the quartz resonator, modes with a frequency of 34 GHz and higher are suitable for measurements (Table 1).

*B. Temperature characteristics of quartz and sapphire resonator based sensors*

As it was mentioned above, the main source of errors in the setup was temperature-related changes in the sapphire resonator parameters. It was not possible to enhance the temperature stabilization in a simple way due to technical reasons. Quartz has lower thermal expansion coefficient and lower temperature dependence of permittivity in comparison with sapphire [7]. In this respect, investigation of resonant frequency shift and changes in inverse quality factor dependencies on temperature has to be performed to analyze the characteristics of quartz and sapphire resonator based sensors. The temperature stabilization system allows to stabilize temperature of the measurement cell in some temperature range above and below room temperature. The resonant frequency and quality factor dependencies on temperature of quartz and sapphire resonator based measurement cells with water filled microfluidic channel were measured in temperature range from 15 to 35°C. The quality factor dependencies on temperature are similar for both measurement cells and therefore they mainly determined by the water parameters, but not by the resonator characteristics. The dependencies of the resonant frequency shift are linear in the mentioned temperature range for both measurement cells. The only difference between sapphire and quartz resonator based sensors is the slope of the dependencies, which is about eight times lower in the case of the quartz resonator based sensor [8]. This fact reflects that the utilization of a quartz resonator instead of a sapphire resonator allows decreasing the temperature dependence of the resonance frequency about eight times.

TABLE I
EXPERIMENTAL SPECTRUM OF THE BARE QUARTZ RESONATOR WITH THE MODES IDENTIFIED BY THE SIMULATIONS

| Mode | Simulation $f$ (GHz) | Simulation $Q$ | Experiment $f$ (GHz) | Experiment $Q$ |
|---|---|---|---|---|
| $HE_{12\,1\,1}$ | 30.475 | 6307 | 30.462 | 6491 |
| $EH_{12\,1\,1}$ | 31.427 | 5227 | 31.456 | 4151 |
| $HE_{13\,1\,1}$ | 32.288 | 13294 | 32.274 | 13354 |
| $EH_{13\,1\,1}$ | 33.385 | 9729 | 33.416 | 8202 |
| $HE_{14\,1\,1}$ | 34.105 | 23888 | 34.089 | 23810 |
| $EH_{14\,1\,1}$ | 35.335 | 15222 | 35.368 | 15086 |
| $HE_{15\,1\,1}$ | 35.926 | 34988 | 35.908 | 34987 |
| $EH_{15\,1\,1}$ | 37.277 | 19903 | 37.311 | 21406 |
| $HE_{16\,1\,1}$ | 37.750 | 42962 | 37.730 | 43058 |
| $EH_{16\,1\,1}$ | 39.212 | 22782 | 39.248 | 25713 |
| $HE_{17\,1\,1}$ | 39.578 | 45519 | 39.556 | 45789 |

*C. Reliability of the simulation*

It was important to check the reliability of the COMSOL model simulation. This was performed by measuring and simulating the cell when filled with liquids with known characteristics. Propanol, ethanol, methanol, and water were selected as reference liquids for this purpose. The permittivity of the reference liquids was obtained from [12] for the appropriate frequency and a temperature of 25°C. The open points in Fig.3 show the dependencies of the measured resonance frequency shift $\Delta f = f_{liq} - f_{air}$ (blue squares) of the air-filled microfluidic channel and changes in inverse quality factor $\Delta(1/Q) = 1/Q_{liq} - 1/Q_{air}$ (red circles) with respect to LUT filled one as a function of the (a) real and (b) imaginary parts of permittivity. The solid points show the values simulated by COMSOL. The good coincidence of the measured and simulated data demonstrates the reliability of the simulation.

*D. Measurement of the glucose in water solutions*

The glucose aqueous solutions were prepared by dissolving glucose powder in bidistilled water. A number of glucose solutions were prepared, with the concentration varying from 0.1% to 4%.

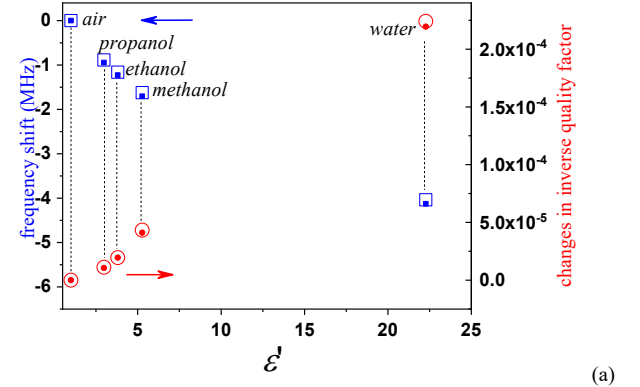

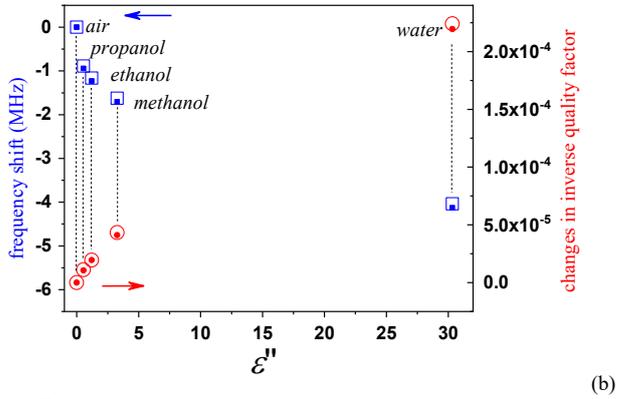

Fig. 3. Dependencies of the experimental (open points) and simulated (solid points) resonant frequency shift $\Delta f = f_{liq} - f_{air}$ (blue rectangles) and changes in inverse quality factor $\Delta(1/Q) = 1/Q_{liq} - 1/Q_{air}$ (red circles) on the (a) real and (b) imaginary parts of the permittivity of reference liquids in the microfluidic channel.

When investigating the glucose in water solutions, it was convenient to measure resonant frequency shift and changes in inverse quality factor with respect to water instead of air. Firstly, most biological liquids are aqueous solutions, and secondly, distilled water was used to clean the channel after each LUT measurement.

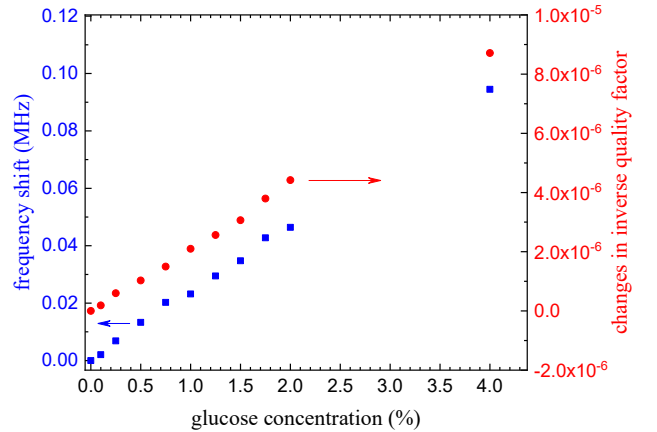

Fig. 4. The dependence of the measured resonance frequency shift $\Delta f = f_{liq} - f_{water}$ (red circles) and changes in inverse quality factor $\Delta(1/Q) = 1/Q_{water} - 1/Q_{liq}$ (blue squares) on glucose concentration.

The measured dependence of the resonance frequency shift $\Delta f = f_{liq} - f_{water}$ (red circles) and changes in inverse quality factor $\Delta(1/Q) = 1/Q_{water} - 1/Q_{liq}$ (blue squares) on glucose concentration is shown in Fig.4. Both values show a monotonic increase with increased concentration of glucose in water.

*Calibration procedure*

The LUT filled in microfluidic channel perturbs the electromagnetic field of the dielectric resonator and causes changes resonance characteristics of the system. The real and imaginary parts of the LUT permittivity change the resonator frequency and quality factor in different ways. In some particular cases of small losses, the real part of LUT permittivity influences only on sensor resonant frequency while introduced losses by a liquid affects only on the inverse quality factor. In these cases, the system can be simply calibrated by a number of reference liquids. As it was shown for the case of sapphire resonator based measurement cell [4] both resonant frequency shift and changes in inverse quality factor depend on the real as well as imaginary parts of LUT permittivity. This explains why it was not possible to use a simple calibration procedure by applying a number of reference liquids. To obtain the complex permittivity of the LUT, the calibration procedure had to be performed via simulation, as shown in [4]. For this calibration, a nomogram chart (Fig.5) was plotted. The values of the resonant frequency shift and changes in inverse quality factor were obtained by simulation, using a varying value for the real part of permittivity with a fixed value for the imaginary part, and vice versa, a varying value for the imaginary part with a fixed value for the real part. The simulated data are represented by open points and connected by solid lines. Solid red points show the experimentally obtained data as functions of glucose concentration in water solutions (0, 0.1, 0.25, 0.5, 0.75, 1, 1.25, 1.5, 1.75, 2, 4 %). The value of the complex water permittivity for frequency equal about 35.7 GHz and temperature about 25°C was taken from [12].

The values of the real and imaginary parts of permittivity can be obtained as shown by dashed lines in Fig.5. In practice, it is difficult to obtain precise values using a graphical plot. To simplify this procedure, a straight line function was constructed using two points. The two simulated points that were closest to the measured point were selected.

*E. Comparison of glucose in water permittivity dependencies obtained by quartz- and sapphire-resonator-based techniques*

The red circles in Fig.6 show the real (solid points) and imaginary (open points) parts of glucose permittivity dependencies on concentration in water solutions as obtained by the quartz-resonator-based measurement cell. The data for the sapphire-resonator-based measurement cell from [4] are also shown in Fig.6, as black rectangles. The line was then moved to cross the measured point and the permittivity value was derived from the resulting straight line equation. It should be noted that, in contrast to the sapphire-resonator-based nomogram chart, the resonant frequency shift has a weak dependence on the imaginary part of LUT permittivity, and changes in inverse quality factor have a weak dependence on the real part of permittivity. Another feature to note is that the closer the permittivity value is to the water permittivity, the weaker these dependencies become. In cases where the permittivity deviates only slightly from the water permittivity value, a simple calibration procedure could be used, although this was not possible in the case of the sapphire-resonator-based measurement cell [4]. However, for the most accurate results, a calibration procedure based on a nomogram chart as described above should be used. The solid and open points correspond to the real and imaginary parts of glucose permittivity, respectively. The data for the real part of permittivity obtained by both measurement cells are in good agreement for concentrations down to 1%.

There is a small shift in the permittivity obtained by the quartz-resonator-based measurement cell in comparison to the sapphire-based cell. This feature is obviously related to the difference in the resonant frequencies of both cells (about 35.4 GHz for the sapphire-based cell and 35.7 GHz for the quartz-based cell). For concentrations of lower than 1% glucose, the quartz-based measurement cell definitely showed greater accuracy in determining the permittivity. For imaginary part permittivity dependencies, the difference between the two cells was lower, since the temperature deviations of resonator characteristics mostly affect the resonant frequency, and do not have a substantial effect on the quality factor.

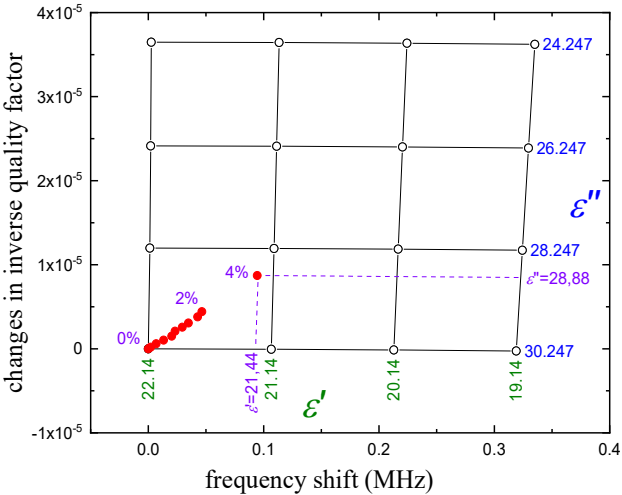

Fig. 5. Nomogram chart. Calculated net (open points) for the resonant frequency shift $\Delta f = f_{liq} - f_{water}$ and changes in the inverse quality factor $\Delta(1/Q) = 1/Q_{water} - 1/Q_{liq}$ for a microfluidic channel filled with substances with varying real and imaginary parts of permittivity (see text for details). Solid points show the experimentally obtained data for resonant frequency shift and changes in inverse factor as functions of glucose concentration in water solutions.

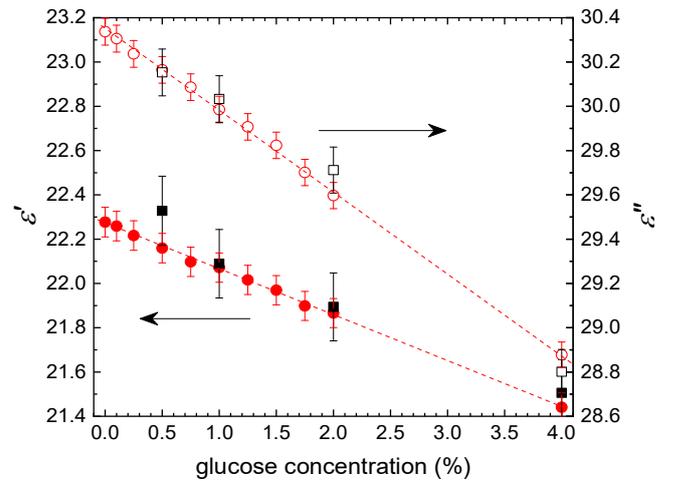

Fig. 6. Dependence of real (solid points) and imaginary (open points) parts of glucose in water permittivity on its concentration in the case of quartz- (circles) and sapphire- (squares) [3] resonator-based measurement cells.

The glucose permittivity detection limit for the quartz-resonator-based measurement cell was lower than 0.1% (about 5.5 mmol/l), which is equal to the concentration of glucose in human blood and about one order of magnitude better than the sapphire-resonator-based measurement cell.

*F. Solution permittivity measurements*

The measured dependencies of the lactalbumin and bovine serum albumin permittivities on concentration in water solutions are shown in Fig.7. The lactalbumin concentration was varied from 250 to 1500 μmol/l, while the bovine serum albumin concentration was varied from 10 to 1000 μmol/l because it has a higher molecular weight. Both substances show linear dependencies of the real and imaginary parts of permittivity on their concentration. As in the case of glucose in water solutions, the substance detection limit was also lower in comparison to the sapphire-resonator-based measurement cell.

The relative accuracy of the permittivity determination of glucose in water solutions using the quartz-resonator-based technique was about 0.7% for the real part and 0.4% for the imaginary part.

The detection limit (DL) of the quartz based sensor is almost one order of magnitude lower in comparison to sapphire resonator based sensor. DL is estimated to be below 0.1% of concentration of low molecular weight substances in water solutions for quartz and between 0.5-1.0% for sapphire based sensor, respectively.

The relative accuracy of the permittivity determination is two times better compared to sapphire based sensor. The accuracy of the real part of permittivity determination of water solutions by quartz-based sensor is equal to 0.7% in comparison with 1.4% for sapphire resonator based sensor. The accuracy of the imaginary part of permittivity determination is equal to 0.7% and 0.4% for sapphire-based and quartz-based sensor, respectively.

The reliability of the method and the simulation model in general is confirmed by measurements of reference liquids such as methanol, ethanol, propanol, water with known parameters as well as by measurement of the glucose permittivity dependence on concentration in water solution and their good correlation with the data, obtained in [4] by sapphire WGM resonator based sensor.

## IV. Conclusion

In this study, an enhanced sensor was developed for determining the microwave complex permittivity of LUT using a quartz WGM resonator with a microfluidic chip. The technique was verified by investigating glucose solutions and comparing the data with results obtained with the sapphire resonator technique. The detection limit obtained for glucose in water solutions was found to be about one order of magnitude lower in comparison to the sapphire-resonator-based measurement cell. The limit obtained is lower than the glucose concentration in human blood. This means that the sensor can be used in studies of biological solutions in the microwave range. The complex permittivity values of lactalbumin and bovine serum albumin water solutions were obtained. The measurements were performed on a single frequency in order to compare the results with the data obtained by the sapphire-resonator-based measurement cell. The measurements can also be performed on a higher-order WGM since its quality factor is even higher. The sensor allows the complex permittivity of small (nanoliter) LUT volumes to be determined with high accuracy.

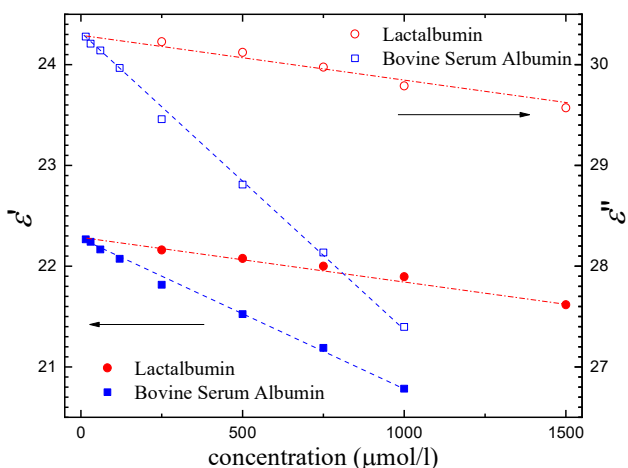

Fig. 7. Dependencies of real (solid points) and imaginary (open points) parts of permittivity on concentration of bovine serum albumin (blue squares and dashed lines) and lactalbumin (red circles and dashed–dotted lines) in aqueous solutions.

References

[1] K. Arkhypova, P. Krasov, A. Fisun, A. Nosatov, V. Lychko, V. Malakhov, "Microwave Dielectrometry as a Tool for the Characterization of Blood Cell Membrane Activity for in vitro Diagnostics," *Int. J. Microwave Wireless Technolog.* vol. 9, No 8, pp. 1569–1574, May 2017. DOI: 10.1017/S1759078717000411

[2] "Keysight Technologies Basics of Measuring the Dielectric Properties of Materials," Application Note 5989-2589EN, Keysight, USA.

[3] Mauro F. Pereira, and Oleksiy Shulika, Ed., THz for CBRN and Explosives Detection and Diagnosis, NATO Science for Peace and Security ser. - B: Physics and Biophysics. Springer, 2017, pp 37-42.

[4] A.I. Gubin, A.A. Barannik, N.T. Cherpak, I.A. Protsenko, A. Offenhaeusser, and S. Vitusevich, "Whispering-gallery-mode resonator technique with microfluidic channel for permittivity measurement of liquids," *IEEE Trans. Microw. Theory Techn.*, vol. 63, No 6, pp. 2003–2009, June 2015. DOI: 10.1109/TMTT.2015.2423289

[5] A.I. Gubin, A.A. Barannik, I.A. Protsenko, N.T. Cherpak, A. Offenhaeusser, and S. Vitusevich, "Biochemical liquids permittivity characterization technique based on whispering-gallery mode resonator with microfluidic channel," in *Proc. of the 43rd European Microwave Conference EuMC'13*, Oct. 6-10 Nuerember, Germany, 2013, pp. 314-317. DOI: 10.23919/EuMC.2013.6686654

[6] A.R.v. Hippel, *Dielectric materials and application.* The M. I. T. Press, Cambridge, Mass. 1954.

[7] D.K. Jha, "Text Book of Heat" in *DPH physics ser.*, Discovery Publishing House, 2004.
[8] A. I. Gubin, I. A. Protsenko, A. A. Barannik, H. Hlukhova, N. T. Cherpak, and S. Vitusevich, "Liquids Microwave Characterization Technique Based on Quartz WGM Resonator with Microfluidic Chip," *Proc. of 48rd Eur. Microwave Conf.*, Sept. 23-28, Madrid, Spain, 2018, pp. 206-209. DOI: 10.23919/EuMC.2018.8541594
[9] COMSOL Multiphysics. [Online]. Available: http://www.comsol.com
[10] Zeonex corporation. [Online]. Available: https://www.zeonex.com
[11] V. N. Skresanov, V. V. Glamazdin, and N. T. Cherpak, "The novel approach to coupled mode parameters recovery from microwave resonator amplitude-frequency response," in *Proc. of the 41rd European Microwave Conference EuMC'11*, Oct. 10-13, Manchester, UK, 2011, pp. 826-829. DOI: 10.23919/EuMC.2011.6101922
[12] J. Barthel and R. Buchner, "High frequency permittivity and its use in the investigation of solution properties," *Pure & App. Chern.*, vol. 63, no. 10, pp. 1473-1482, Oct. 1991. DOI: 10.1351/pac199163101473

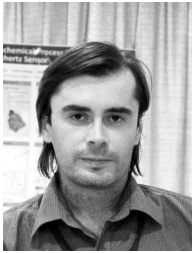

**Alexey I. Gubin** was born in Kharkiv, Ukraine, on March 29, 1978. He received a diploma in radiophysics and electronics from V.N. Karazin Kharkov National University, Kharkiv, Ukraine in 2000 and a Ph.D. degree (physical and mathematical sciences) in radiophysics from O. Ya. Usikov Institute for Radiophysics and Electronics, National Academy of Science of Ukraine, Kharkiv, Ukraine, in 2007.

Since 2000, he has been working at O. Ya. Usikov Institute for Radiophysics and Electronics, National Academy of Science of Ukraine, Kharkiv, Ukraine, currently in the role of Senior Researcher. In 2017 he became Associate Professor. He is co-author of more than 70 scientific publications. His scientific activities encompass the study of the microwave to submillimeter-wave characteristics of condensed matter using techniques based on non-resonant grazing incidence reflectivity technique and whispering-gallery-mode dielectric resonator based technique.

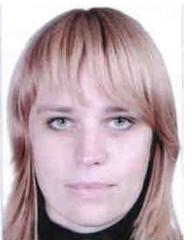

**Irina A. Protsenko** received a diploma in radio engineering from Kharkiv National University of Radio Electronics in 2005 and a Ph.D. degree (physical and mathematical sciences) in radiophysics from O. Ya. Usikov Institute for Radiophysics and Electronics, National Academy of Science of Ukraine, Kharkiv, Ukraine, in 2018.

She has been working at the Institute of Radiophysics and Electronics, National Academy of Science of Ukraine, Kharkiv, Ukraine since 2005. Her current research interests include the measurement of the electrophysical parameters of materials.

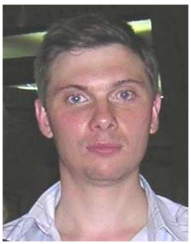

**Alexander A. Barannik** was born in Kharkiv, Ukraine, on April 23, 1975. He received a diploma in cryogenic techniques and technology from Kharkiv State Polytechnic University, Kharkiv, Ukraine, in 2000 and a Ph.D. degree (physical and mathematical sciences) in radiophysics from Kharkiv National University of Radio Electronics, Kharkiv, Ukraine in 2004.

Since 2000 he has been working at the O. Ya. Usikov Institute for Radiophysics and Electronics, National Academy of Science of Ukraine, Kharkiv, Ukraine, as a Junior Researcher (2000), Scientific Researcher (2005), and Senior Researcher (2007). He has co-authored over 100 scientific publications and he has five patents. His scientific activities focus on the study of the microwave characteristics of condensed matter including HTS, dielectrics, and liquids using whispering-gallery-mode dielectric resonators. He also studies the electrodynamic properties of various types of whispering-gallery-mode resonators.

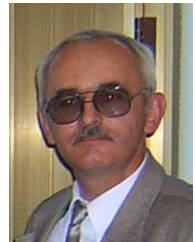

**Alexandr A. Lavrinovich** received a diploma in electronics devices from Kharkov National University of Radio Electronics, Ukraine, in 1978 and a Ph.D. degree (physical and mathematical sciences) from O. Ya. Usikov Institute for Radiophysics and Electronics, National Academy of Science of Ukraine, Kharkiv, Ukraine, in 1997.

Since 1978 (excepting 1980–1982, when he served as an officer in the army), he has been working at the O. Ya. Usikov Institute for Radiophysics and Electronics, National Academy of Science of Ukraine, Kharkiv, Ukraine, currently in the role of Senior Researcher. In 2000 he became Associate Professor. He has authored and co-authored about 100 scientific publications and he has nine patents. His scientific activities encompass the study of quantum radio physics (low-noise amplifiers – masers), magnetic systems, and the microwave characteristics of solid-state matters and liquids including HTS and others, using waveguide structures and whispering-gallery-mode resonators.

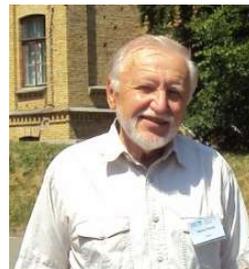

**Nickolay T. Cherpak** received PhD and D.Sc. degrees in 1970 and 1987, respectively.

He is Leading Researcher at O. Ya. Usikov Institute for Radiophysics and Electronics, National Academy of Sciences of Ukraine, Kharkiv, Ukraine. He was previously engaged in the development of low-noise solid-state microwave and millimeter-wave maser amplifiers. At present he is the head of a research group that is developing a millimeter-wave measurement technique for the characterization of condensed matter such as new, unconventional superconductors and lossy dielectric liquids. The group also performs the relevant physical studies.

In 1999, Prof. Cherpak received the I. Puljuy Award of Presidium of NASU for the best experimental work on RF and microwave studies of HTS. He also received the Academician Sinelnikov Award from the Kharkiv Regional Administration in 2013 and the L. Shubnikov Award from the Presidium of NASU for the best experimental work on microwave impedance studies of new, unconventional superconductors in 2016. Prof. Cherpak is the author and co-author of about 350 publications, including three books, five reviews, and over 20

FSU, Ukraine, and USA patents, and has also supervised seven Ph.D. dissertations. He is also a Professor at the National Technical University (NTU) – Kharkov Polytechnic Institute. He is an editorial staff member of two scientific journals. Prof. Cherpak is a member of URSI Commission E in Ukraine, Chairman of the IEEE Joint East Ukraine Chapter, a EuMA member, and served as a delegate for Ukraine at the EuMA GA. He has been chairman, program committee member, and session chair of a number of international conferences and workshops.

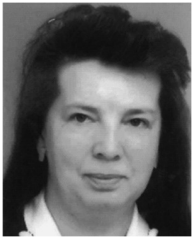

**Svetlana Vitusevich** received an M.Sc. degree in radiophysics and electronics from Kiev State University, Kiev, Ukraine, in 1981, a Ph.D. degree in physics and mathematics from the Institute of Semiconductor Physics (ISP), Kiev, Ukraine, in 1991, and a Dr. habil. degree from the Supreme Attestation Commission of the Ukraine, Kiev, Ukraine, and Technical University of Dortmund, Dortmund, Germany, in 2006.

From 1981 to 1997, she was with the ISP, as Researcher (1981), Scientific Researcher (1992), and Senior Scientific Researcher (1994). From 1997 to 1999, she worked at the Institute of Thin Films and Interfaces (ISI), Forschungszentrum Jülich (FZJ), Jülich, Germany, as an Alexander von Humboldt Research Fellow. Since 1999, she has been a Senior Scientific Researcher with the Institute of Complex Systems – Bioelectronics (ICS-8, formerly Peter Grünberg Institute – PGI-8, IBN-2, ISG-2, and ISI), FZJ. In 2006, she became the leader of a working group whose main topic related to 1/f noise in quantum heterostructures and its up-conversion into phase noise in high-frequency oscillators, including high-quality resonators. She is an active teacher with the Technical University of Dortmund, Dortmund, Germany. She has authored or co-authored over 170 papers in refereed scientific journals. She holds ten patents, among them a dual-mode resonator for simultaneous measurement of dielectric relaxation and electrical conductivity of liquids in closed bottles. Her research interests have focused on transport and noise properties of different kinds of materials for development of label-free biosensors and novel device structures for future information technologies.